\documentclass[reprint,
superscriptaddress,groupedaddress,
showpacs,
nobibnotes,
amsmath,amssymb,
aip,
apl,
floatfix,
showkeys,
]{revtex4-1}

\usepackage[utf8]{inputenc}
\usepackage[T1]{fontenc}
\usepackage[english]{babel}
\usepackage{cancel}
\usepackage{graphics}
\usepackage{amsmath,amsfonts,bm,mathrsfs,dsfont}
\usepackage{amssymb}
\usepackage{bigints}
\usepackage{color}
\usepackage{hyperref}
\usepackage{comment}

\usepackage{booktabs}

\usepackage[oldvoltagedirection]{circuitikz}
\usepackage{tikz}
\usetikzlibrary{patterns,arrows,shapes,arrows,intersections,decorations}

\renewcommand{\vec}[1]{\mathbf{#1}}

\begin{document}

\title{Hydrodynamical study of Terahertz emission in magnetized graphene field-effect transistors}

\author{Pedro Cosme}
\email{pedro.cosme.e.silva@tecnico.ulisboa.pt} 
\affiliation{Instituto Superior T\'ecnico, 1049-001 Lisboa, Portugal}%
\affiliation{Instituto de Plasmas e Fus\~ao Nuclear, 1049-001 Lisboa, Portugal}

\author{Hugo Ter\c{c}as}
\email{hugo.tercas@tecnico.ulisboa.pt} 
\affiliation{Instituto Superior T\'ecnico, 1049-001 Lisboa, Portugal}
\affiliation{Instituto de Plasmas e Fus\~ao Nuclear, 1049-001 Lisboa, Portugal}

\begin{abstract}
Several hydrodynamic descriptions of charge transport in graphene have been presented in the late years. We discuss a general hydrodynamic model governing the dynamics of a two-dimensional electron gas in a magnetized field-effect transistor in the slow drift regime. The Dyakonov--Shur instability is investigated, including the effect of weak magnetic fields (i.e. away from Landau levels). We show that the gap on the dispersion relation prevents the instability from reaching the lower frequencies, thus imposing a limit on the Mach number of the electronic flow. Furthermore, we discuss that the presence of the external magnetic field decreases the growth rate of the instability, as well as the saturation amplitude. The numerical results from our simulations and the presented higher order dynamic mode decomposition support such reasoning.
\end{abstract}

\maketitle

\section{Introduction}

In recent years, the scientific community has witnessed the emergence of integrated-circuit technology with bi-dimensional (2D) materials. In this scope, graphene is undoubtedly one of the most prominent materials. Among the many applications of graphene, the possibility of resorting to plasmonics instabilities to trigger the emission, or conversely, the detection, of THz radiation has been an active field of study \cite{Hosseininejad2018,Mendl2019,Wigger2017,Otsuji2014}. 
The explored mechanisms for the creation and control of plasmons in graphene commonly rely on graphene field-effect transistors (GFET), which allow to control the Fermi level while being easily combined in integrated circuitry.

One of the defining characteristics of graphene is its high electron mobility, as a consequence of the weak scattering between electrons and phonons, defects, or impurities, which leads to large electron--impurity mean free path $\ell_\text{imp}$. Indeed, ultra-clean samples of graphene encapsulated by hexagonal boron nitride (hBN) \cite{Son2018} or hBN--graphene--WSe$_2$ structures \cite{Banszerus2019} exhibit a mobility $\mu>3.5\times 10^{5}\,\mathrm{cm^2V^{-1}s^{-1}}$. Yet, the electron--electron scattering is significant, resulting in a short mean free path $\ell_{ee}$ at room temperature. Thereby, it is possible to design a system of size $L$ under the condition $\ell_{ee}\ll L \ll \ell_\text{imp}$. In such a regime, the collective behavior of carriers can be accurately described hydrodynamically \cite{Chaves2017,Narozhny2017,Svintsov2013,Mendl2019,Lucas2018,Muller2009}, with some recent experimental results validating this approach \cite{Sulpizio2019,Mayzel2019,Berdyugin2019MeasuringFluid}. 

Given the massless nature of graphene electrons, a relativistic description is required for velocities near the Fermi velocity $v_F$. However, for the usual operation conditions of GFETs, the velocity of the carriers is expected to saturate far below $v_F$ \cite{Wilmart2020,Yamoah2017,Dorgan2010b}. As such, we here model graphene plasmons making use of a hydrodynamic set of equations valid in the regime $v\ll v_F$. Moreover, we operate at room temperature, such that the Fermi level is large enough to prevent interband transitions,  $E_F\gg k_BT$. 

The Dyakonov--Shur (DS) instability has been extensively studied for high-mobility semi-conductors as a mechanism for emission/detection of THz radiation \cite{Dyakonov1993,Crowne2000DyakonovShurTransistor} and has recently been considered in graphene devices \cite{Cosme2020, Lucas2018c}. However, few works have approached the issue under the influence of magnetic fields \cite{Dyakonova2005MagneticTransistors,Kushwaha2001InfluenceTransistor,Zhang2013}. In this work, we investigate the DS instability taking place in GFETs in the regime of weak magnetic fields, i.e. away from the Landau levels. Due to the appearance of a gap, the difference of frequency between the forward and backward plasmon modes is decreased, leading to an attenuation of the DS frequency and growth rate. We also show that the emergence of a transverse (Hall) current in the channels in the nonlinear regime is responsible for the decreasing of the electron saturation amplitude. 

\section{Hydrodynamic Model for Graphene Electrons}

\begin{figure}[t!]
    \centering
    \includegraphics[width=0.8\linewidth]{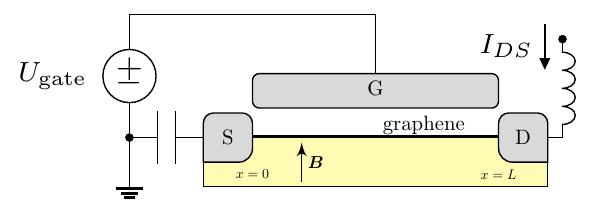}
    
    \caption{Schematic representation of a graphene channel field-effect transistor with a top gate (G). The presented setup also shows the Dyakonov--Shur impedance realization at source (S) and drain (D). The magnetic field is perpendicular to the channel.}
    \label{fig:gfet}
\end{figure}

The fact that the electrons in graphene behave as massless Dirac fermions poses the major difficulty for the development of hydrodynamic models: not only do carriers have zero mass, but also the effective inertial mass tensor diverges \cite{Ashcroft1976}. A naive approach would dictate to define an effective mass as    
\begin{equation}
    m^\star=\frac{\hbar k_F}{v_F}=\frac{\hbar\sqrt{\pi n}}{v_F},
    \label{eq:DrudeMass}
\end{equation}
where $\hbar k_F$ is the Fermi momentum and $n$ is the electron 2D number density. This definition is extensively used in the literature \cite{Chaves2017,Lucas2018,Svintsov2013}, and recent developments based on quantum kinetic theory propose corrections to it \cite{Figueiredo2020WignerPlasmas}. Since the electronic fluid is compressible, the effective mass is not a conserved quantity, contrary to customary fluids. For typical conditions in GFETs, the effective mass is expected in the range

\begin{equation}
    2.7\,\mathrm{keV/c^2}\ll m^\star\ll 270\,\mathrm{keV/c^2},
\end{equation}
lying fairly below the free electron mass. 

Starting from the Boltzmann equation for the distribution function $f=f(\bm{r},\bm{p},t)$ 
\begin{equation}
    \frac{\partial}{\partial t} f + v_F\mathbf{\frac{\bm p}{\vert \bm p\vert}}\cdot\bm\nabla_{\bm r}f +  \mathbf{F}\cdot\bm\nabla_{\bm p} f = \widehat{\mathcal{C}}[f],\label{eq:Boltzmann}
\end{equation}
one can derive the hydrodynamic model for electronic transport in graphene. Here, the collision operator can be taken in the Bhatnagar--Gross--Krook approximation \cite{Rieutord2015,Haas2011}, $\widehat{\mathcal{C}}[f]=(f_{\text{Equilibrium}}-f)/\tau$. However, since we are interested in mesoscopic effects with small Knudsen number, $v \tau/ L \ll 1 $, and time scales much longer than the collision time, we can safely set $\widehat{\mathcal{C}}[f]\approx 0$. This does not imply the absence of electron-electron collisions in the electronic fluid, but rather that they occur fast enough to maintain the local equilibrium. 

By integrating the zero-order momentum of Eq. \eqref{eq:Boltzmann}, yields the continuity equation 
\begin{equation}
    \frac{\partial n}{\partial t} +\bm{\nabla}\cdot\left( n\mathbf{v}\right) = 0.
    \label{continu}
\end{equation}
Furthermore, the first momentum of Eq. \eqref{eq:Boltzmann} leads to
\begin{equation}
    \frac{\partial \vec{v}}{\partial t} %
+ \frac{(\vec{v}\cdot \bm{\nabla} )\vec{v}}{2}
+\frac{1}{n m^\star}\bm{\nabla}\cdot\mathbb{P}-\frac{\vec{F}}{m^\star}=0,\label{preeuler}
\end{equation}
where $\mathbb{P}$ is the pressure stress tensor and $\mathbf{F}$ the resultant external force. As we can see, the variation of the effective mass introduces a $1/2$ factor to the convective term. Such correction breaks the Galilean invariance of the system, leading to an unusual expression for the dispersion relation in the presence of a Doppler shift \cite{Cosme2020}.  

The \emph{hydrostatic} diagonal terms of the pressure, $\mathbb{P}=P\delta_{ij}$, is given by the 2D Fermi-Dirac pressure \cite{Landau1980a,Giuliani2005QuantumLiquid,Chaves2017} 
\begin{equation}
P=\frac{2(k_BT)^3}{\pi \hbar^2v_F^2}\,\mathfrak{F}_2\left(\frac{E_F}{k_BT}\right),
\end{equation}
where $\mathfrak{F}_2$ is the complete Fermi-Dirac pressure, which at room temperature, $E_F\gg k_BT$, gives
\begin{equation}
P=\frac{E_F^3}{3\pi(\hbar v_F)^2}+\mathcal{O}\left(\frac{k_BT}{E_F}\right)^2\simeq\frac{\hbar v_F}{3 \pi}\big(\pi n\big)^{\frac{3}{2}}. \label{eq:FermiLiquidPress}
\end{equation}
As such, the pressure term in \eqref{preeuler} reduces to 
\begin{equation}
    \frac{1}{nm^\star}\bm{\nabla} P=\frac{v_F^2}{2n}\bm{\nabla} n.
\end{equation}

The off-diagonal elements of the pressure in Eq. \eqref{preeuler} describe the viscous terms of the fluid. The kinematic viscosity near the Dirac point is $\nu\simeq v_F\ell_{ee}/4\sim 2.5\!\times\!10^{-3}\,\mathrm{m^2s^{-1}}$; however, at room temperature $T\ll T_F$ this value increases to $\nu\sim 0.1\,\mathrm{m^2s^{-1}}$ \cite{Narozhny2019MagnetohydrodynamicsViscosities,Lucas2018,Muller2009,Torre2015a,Levitov2016ElectronGraphene}, and the corresponding Reynolds number of the electron fluid is 
\begin{equation}
    \mathrm{Re}\sim\frac{Lv_0}{0.1 \mathrm{m^2s^{-1}}}.  
\end{equation}
A suitable choice of the system parameters can be made such that ${\rm Re}\gg 1$, rendering the viscous effects negligible. As a matter of fact, our simulations performed for moderate values of the Reynolds number have not shown any significant difference from the inviscid case, apart from the expected suppression of higher frequency content and subsequent smoothing of the waveforms.
 \par
 
For a magnetized graphene electron gas in the field-effect transistor configuration, as depicted in Fig. \ref{fig:gfet}, the force term results from the combined effect of the gate and the cyclotron (Lorentz) force,

\begin{equation}
\vec{F}=-\bm \nabla U_{\rm gate} -\frac{e}{m^\star}\vec{v}\times\vec{B},    
\end{equation}
where $U_{\rm gate}$ is the gate voltage,

\begin{equation}
    U_{\text{gate}}=en\left(\frac{1}{C_g}+\frac{1}{C_q}\right),
    \label{eq:Ugate}
\end{equation}
with $C_g$ and $C_q$ denoting the geometric and the quantum capacitances \cite{Zhu2009a,Sarma2010}. For typical carrier densities $n\gtrsim10^{12}\,\mathrm{cm}^{-2}$, the quantum capacity dominates, $C_q\gg C_g$, and $U_{\rm gate}\simeq en/C_g=end_0/\epsilon$.  

\subsection{Enhanced diamagnetic drift}

In the presence of a magnetic field, the system is subject to Lorentz force and, taking the steady state of eq. \eqref{preeuler} leads to
\begin{equation}
\frac{v_F^2}{2n}\bm{\nabla}n
+\frac{s^2}{\sqrt{n_0n}}\bm{\nabla}n+\frac{e\vec{v}\times\vec{B}}{m^\star}=0,
\end{equation}
where $s=\left(e^2 d v_F\sqrt{ n_0}/\varepsilon \hbar\sqrt{\pi}\right)^{1/2}$ is the screened plasmon sound velocity. The drift velocity perpendicular to $\vec{B}$ can be retrieved as 
\begin{equation}
    \vec{v}_\perp=\frac{\underline{S}^2m^\star}{n_0e}\frac{\bm{\nabla}n\times\vec{B}}{\vec{B}^2},
\end{equation}
with $\underline{S}^2=s^2+v_F^2/2$ which is analogous to a diamagnetic drift\cite{Chen2016IntroductionFusion} in plasmas. Here, however, the drift is not only due to the pressure gradient\cite{Chen2016IntroductionFusion} but has the added contribution of the force drift since $\vec{F}\sim\bm{\nabla}n$ as well. Thus, the fluid has a larger diamagnetic drift compared to what would be expected from the pressure itself. In the case of wave or shock propagation along the GFET channel, as the density gradient will be mostly in the $x$ direction and, therefore, the diamagnetic drift will give rise to a transverse Hall current. 

\subsection{Magneto-plasmons in graphene FETs}

\begin{figure}[!t]
    \centering
    \includegraphics[width=0.90\linewidth]{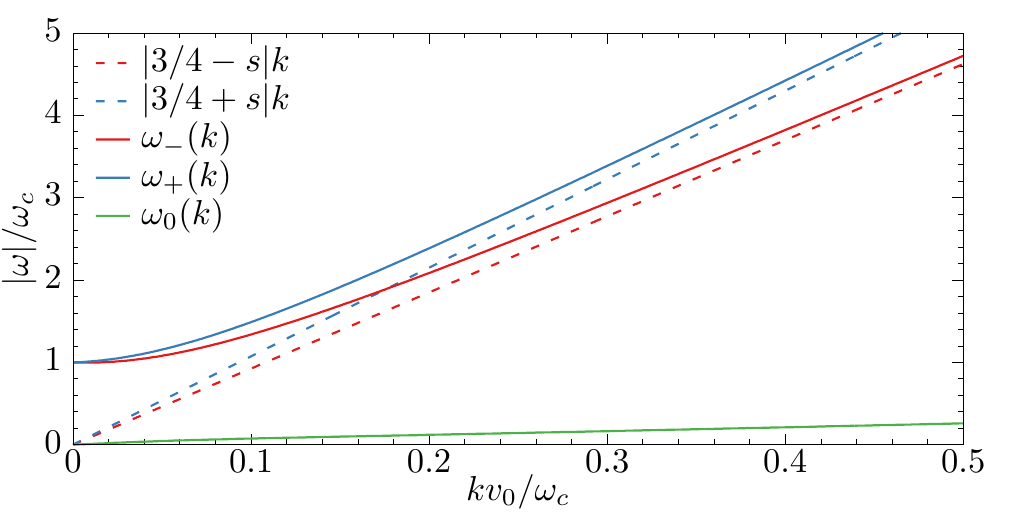}
    \caption{Magneto-plasmon dispersion in graphene FETs. Solutions of the dispersion relation in EQ. \eqref{eq:secularB} with  $\underline{S}/v_0=10$ (solid lines) alongside the solutions in the absence of magnetic field (dashed lines).}
    \label{fig:Bdisper}
\end{figure}

Considering an uniform field $\vec{B}=B_0\bm{\hat{z}}$ perpendicular to the graphene layer and writing $\vec{v}=v_x\bm{\hat{x}}+v_y\bm{\hat{y}}$ while looking for propagation along $x$, $\vec{k}=k\bm{\hat{x}}$, linearization of Eqs. \eqref{continu} and \eqref{preeuler}, with $\vec{v}=(v_0+v_x)\bm{\hat{x}}+v_y\bm{\hat{y}}$ and $n=n_0+n_1$, reads in Fourier space
\begin{subequations}
\begin{gather}
\left(\omega -kv_0\right)\tilde{n}_1=kn_0\tilde{v}_{x}, \\
\left(\omega - \frac{kv_0}{2}\right)\tilde{v}_x =k\frac{\underline{S}^2}{n_0}\tilde{n}_1-i\omega_c\tilde{v}_y, \\
\left(\omega - \frac{kv_0}{2}\right)\tilde{v}_y=i\omega_c\tilde{v}_x,
\end{gather}\label{eq:fourier1}%
\end{subequations}
where $\omega_c=eB/m^\star$ is the cyclotron frequency. Note that as the effective mass is much smaller than the electron mass, $m^\star\ll m_e$, it is possible to access high cyclotron frequencies with modest fields; for a typical excess density of $10^{12}\,\mathrm{cm}^{-2}$ $\omega_c/B=9\,\mathrm{THzT^{-1}}$. Furthermore, combining \eqref{eq:fourier1} yields the relation
\begin{multline}
\left(\omega-kv_0\right)\left[\left(\omega-\frac{kv_0}{2}\right)^2\!\!-\omega_c^2\right]\!\!=\underline{S}^2k^2\left(\omega-\frac{kv_0}{2}\right).\label{eq:secularB}
\end{multline}
With this dispersion relation, the propagating solutions $\omega_\pm(k)$ coalesce to $\omega_c$ as $k\!\!\rightarrow\!\!0$, opening a gap at the origin as patent in Fig. \ref{fig:Bdisper}, whereas for large $k$ we recover the unperturbed solutions $\omega\simeq (3/4v_0\pm\underline{S})k$. Moreover, a third solution $\omega_0(k)\simeq kv_0/2$ is also present.

\section{Dyakonov--Shur Instability}

\begin{figure}[!t]
    \centering
    \includegraphics[width=0.90\linewidth]{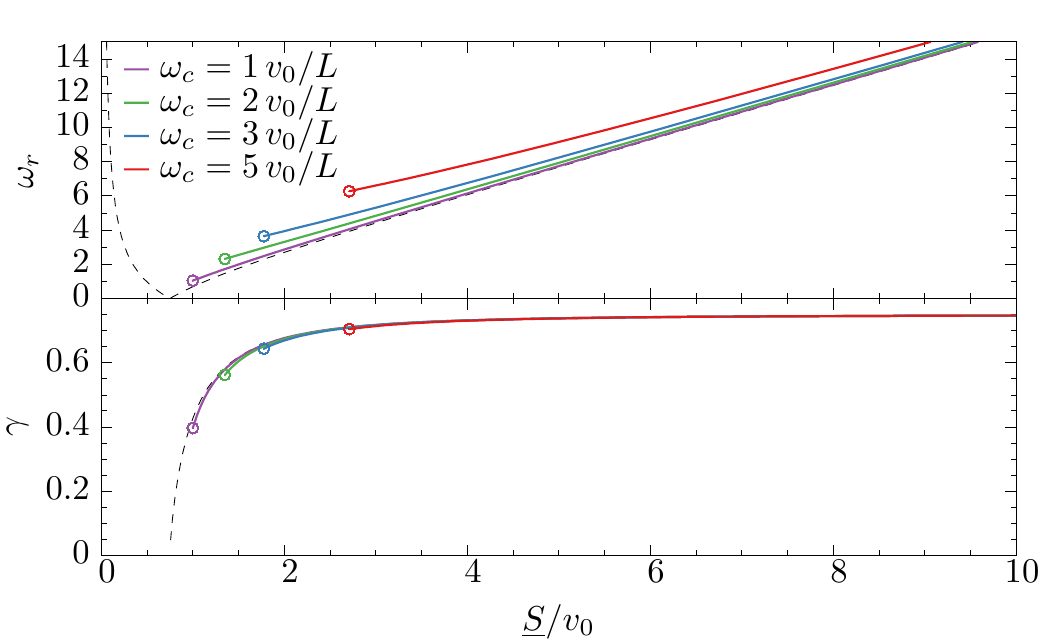}
    \caption{Numerical solutions for frequency and growth rate (in units of $v_0/L$) of Dyakonov--Shur instability for several cyclotron frequencies $\omega_c$ (coloured dots) and analytical solution \eqref{eq:omegatotal} corresponding to $B=0$ (dashed black line). Although there is no significant change in the real part of the frequency, the growth rate diminishes slightly.}
    \label{fig:efeitosB}
\end{figure}

The hydrodynamic model in Eqs. \eqref{continu} and \eqref{preeuler} contains an instability under the boundary conditions of fixed density at the source $n(x=0)=n_0$ and fixed current density at the drain $n(x=L)v(x=L)=n_0v_0$, dubbed in the literature as the Dyakonov--Shur (DS) instability \cite{Dyakonov1993,Dyakonov2011}. The latter arises from the multiple reflections of the plasma waves at the boundaries, which provide positive feedback for the incoming waves driven by the current at the drain.
From an electronic point of view, the peculiar boundary conditions correspond to an AC short circuit at the source, forcing the voltage (and so the carriers density) to remain constant, and an AC open circuit at the drain setting the current constant \cite{Crowne1997,Barut2019}. Thus, these conditions can be implemented with a low-reactance capacitor on the source and a high-reactance inductor on the drain \cite{Fay2020}, as outlined in Figure \ref{fig:gfet}.

The asymmetric boundary conditions described above imply that the counterpropagating wave vectors need to comply with the relation 
\begin{equation}
    \frac{k_+}{k_-}=e^{i(k_+-k_-)L},  \label{eq:DScondition}
\end{equation}
where 
\begin{equation}
k_{\pm}=\frac{\frac{3}{4}\omega \mp \text{Sgn}(\omega)\sqrt {s^2\left (\omega^2 - \omega_c^2 \right) + \left (\frac {3}{4}\omega_c \right)^2}}{\left (\frac{3}{4}\right)^2 - s^2}
.
\label{eq:vectors}
\end{equation}
This condition leads to complex solutions, $\omega=\omega_r+i\gamma$, where $\omega_r$ is the electron oscillation frequency and $\gamma$ is the instability growth rate \cite{Dyakonov1993,Dmitriev1997,Crowne1997}. Numerical inspection of Eq. \eqref{eq:DScondition} provides the results depicted in Fig. \ref{fig:efeitosB}. In the unmagnetized case, the instability condition can be analitically solved
\begin{equation}
\begin{gathered}
    \omega_r = \frac{|\underline{S}^2-\left(\frac{3}{4}v_0\right)^2|}{2L\underline{S}}\pi, \\
    \gamma = \frac{\underline{S}^2-\left(\frac{3}{4}v_0\right)^2}{2L\underline{S}}\log\left|\frac{\underline{S}+\frac{3}{4}v_0}{\underline{S}-\frac{3}{4}v_0}\right|.
\label{eq:omegatotal}
\end{gathered}
\end{equation}
Plasmonic dynamical instability takes place for $S/v_0>3/4$, i.e. in the \emph{subsonic} regime. The fact that the instability develops in such a regime is advantageous from the technological point of view, as it allows the operation of the GFET far from the velocity saturation\cite{Schwierz2010a,Wilmart2020}.  
Moreover, when $S\gg v_0$ the frequency is dominated by the $S/L$ ration as $\omega_r\sim\pi\underline{S}/2L$ while $\gamma\sim 3v_0/4L$. Then, given the dependence of $S$ with gate voltage, and as $v_0n_0\sim I_{\rm DS}/We$, with $I_{\rm DS}$ representing the source-to-drain current and $W$ the transverse width of the sheet, the frequency can be tuned by the gate voltage and injected drain current, not being solely restricted to the geometric factors of the GFET. 
 
In the presence of the magnetic field, the solutions of \eqref{eq:DScondition}
reveal that the growth rate of the instability decreases slightly, which is more evident around the transonic regime, while at the subsonic case the influence of the magnetic field on the growth rate is less noticeable (Fig. \ref{fig:efeitosB}). This observation contradicts what has been previously reported in Ref. \cite{Zhang2013}. Regarding the frequency, the magnetic field introduces a small shift from the unmagnetized scenario.

The reason for our results to differ from those presented in \cite{Zhang2013} lies in the treatment of the wave vector solutions. In the cited work the cyclotron frequency $\omega_c$ is a priori normalized to $\underline{S}/L$. Such approach simplifies the problem as it artificially linearises \eqref{eq:vectors}. However, this obscures the analysis as in a $\omega$ vs. $\underline{S}$ plot, the cyclotron frequency would also be varying. Moreover, the gap of the dispersion relation opened by the magnetic field suppresses frequencies below $\omega_c$; hence, as one approaches the sonic regime $\underline{S}\sim v_0$, the real part of the frequency drops and reaches the cut-off. Thus, leaving the solutions on Fig.\ref{fig:efeitosB} with an endpoint.

\section{Numerical Simulation}

\begin{figure}[!t]
    \centering
    \includegraphics[width=0.95\linewidth]{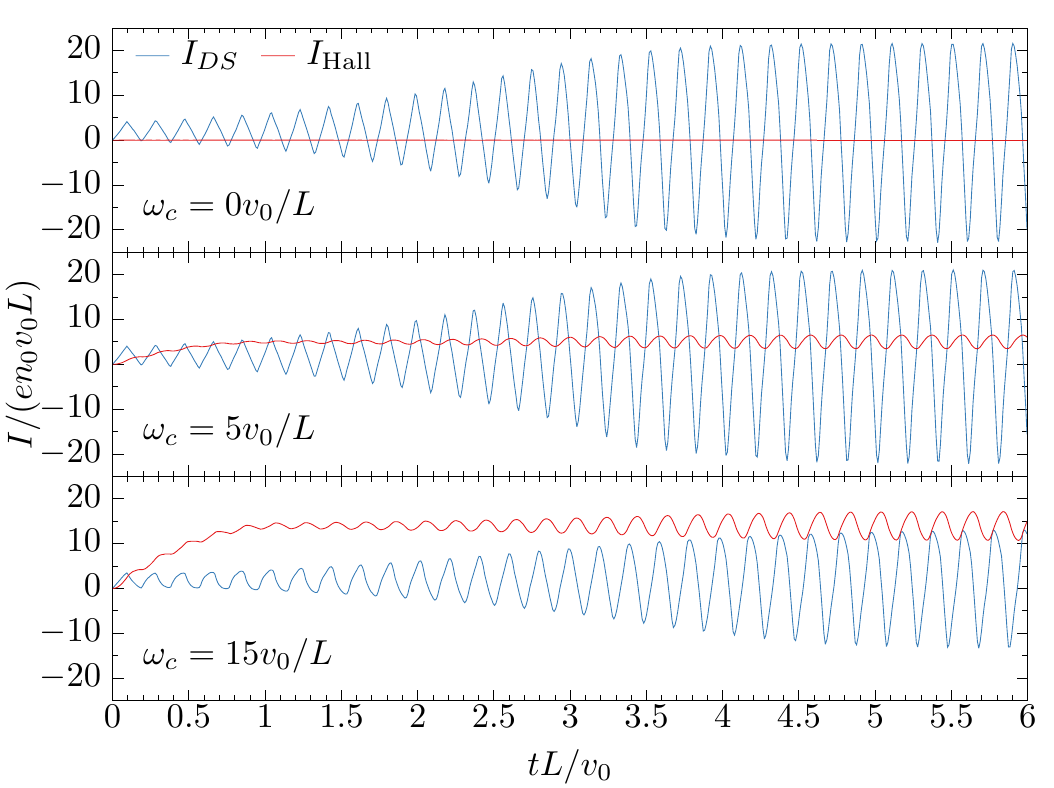}
    \caption{Evolution of drain-to-source and Hall currents across the graphene channel for distinct values of cyclotron frequency. The presence of magnetic filed diverts part of the current to the transverse direction and diminishes the growth rate of instability. All three simulations performed with $S=20v_0$ and $v_F=10v_0$.}
    \label{fig:simul}
\end{figure}

\begin{figure}[!b]
    \centering
    \includegraphics[width=0.90\linewidth]{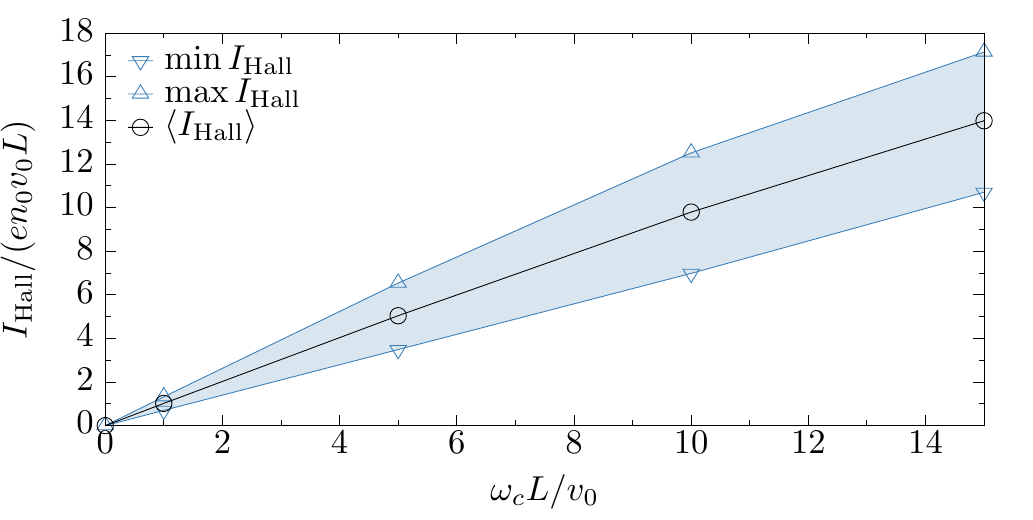}
    \caption{Hall current response with the applied magnetic field. All simulations performed with $S=20v_0$ and $v_F=10v_0$. }
    \label{fig:Ihall}
\end{figure}
In order to perform the simulations revealing the late-stage (nonlinear) evolution of the plasmon wave in the FET channel, the hydrodynamical equations have been recast into a conservation form plus a magnetic source term. Resorting to the mass flux density $\vec{p}=m^\star n\vec{v}$, the continuity and momentum equation can be written in the equivalent form   
\begin{subequations}
\begin{equation}
\frac{\partial n}{\partial t}+\bm{\nabla}\!\cdot\!\frac{\vec{p}}{\sqrt{n}}=0,    
\end{equation}
\begin{multline}
\frac{\partial \vec{p}}{\partial t}+\bm{\nabla}\!\cdot\!\left( \frac{ \vec{p}\otimes \vec{p}}{{n}^{3/2}} + \frac{v_F^2}{v_0^2}\frac{n^{3/2}}{3}\mathds{1}+\frac{S^2}{v_0^2}\frac{{n}^2}{2}\mathds{1}\right)+\\+\frac{\omega_c}{\omega_0}\frac{\vec{p}\times\vec{\hat{z}}}{\sqrt{n}}=0.
\end{multline}
\label{eq:sistema_numerico}
\end{subequations}
This hyperbolic system of differential equations has been solved with a finite volume Lax-Wendroff method \cite{Hirsch2007,LeVeque1992}, the two-step Richtmyer scheme for nonlinear systems \cite{LeVeque1992}. The simulation of system \eqref{eq:sistema_numerico}, as well as the computation of the observable electronic quantities of the GFET, has been carried with a software specifically developed for the task \cite{Cosme2020TETHYSSimulation.}. Our simulations confirm that the magnetic field reduces the instability growth rate, as expected for the subsonic regime (Fig.\ref{fig:efeitosB}). The average value and oscillation amplitude of the quantities along the channel are also reduced (Tab.\ref{tab:tabela_ids}), as the diamagnetic current removes a fraction of the electrons participating in the longitudinal oscillation. A typical situation for the current density at source can be seen in Fig.\ref{fig:simul}. The latter reveals that the magnetic drift is responsible for a transverse current, which could be exploited for a directional coupler operating in the THz regime \cite{He2014Graphene-basedCoupler}. In the present case, we are dealing with plasmons, but it may also be applicable to the case of surface-plasmon polaritons \cite{Hwang2019DirectionalResonators}. Indeed the applied magnetic field can control not only the average $I_\text{Hall}$ value but also amplify the amplitude of its oscillation as patent on Fig.\ref{fig:Ihall}.

\begin{table}[!t]
    \centering
        \caption{Average values and extrema of the drain-to-source and Hall currents (in units of $en_0v_0L$) at the nonlinear regime with the imposition of a cyclotron frequency $\omega_c$ (in units of $v_0/L$). All simulations were performed with $S=20v_0$ and $v_F=10v_0$.}
    \label{tab:tabela_ids}
\begin{tabular*}{\linewidth}{l @{\extracolsep{\fill}} ccc ccc}
   \toprule
        $\omega_c$& $\langle I_\text{Hall}\rangle$ & $\min I_\text{Hall}$ & $\max I_\text{Hall}$  & $\langle I_{DS}\rangle$ & $\min I_{DS}$ & $\max I_{DS}$ \\
    \midrule    
        $0$ & --- & --- & --- & $2.053$ & $-22.884$ & $21.584$  \\
        $1$ & $\phantom{1}1.017$ & $\phantom{1}0.701$ & $\phantom{1}1.322$ & $2.051$ & $-22.851$ & $21.557$\\
        $5$ & $\phantom{1}5.039$ & $\phantom{1}3.486$ & $\phantom{1}6.539$ & $2.042$ & $-22.183$  & $21.037$\\        
        $10$& $\phantom{1}9.796$ & $\phantom{1}6.979$ & $12.507$ & $1.971$ & $-19.053$& $18.835$  \\
        $15$& $13.979$ & $10.703$ & $17.134$ & $1.734$ & $-13.356$ & $13.029$ \\        
    \bottomrule 
\end{tabular*}
\end{table}

\begin{figure}[!t]
    \centering
    \includegraphics[width=0.95\linewidth]{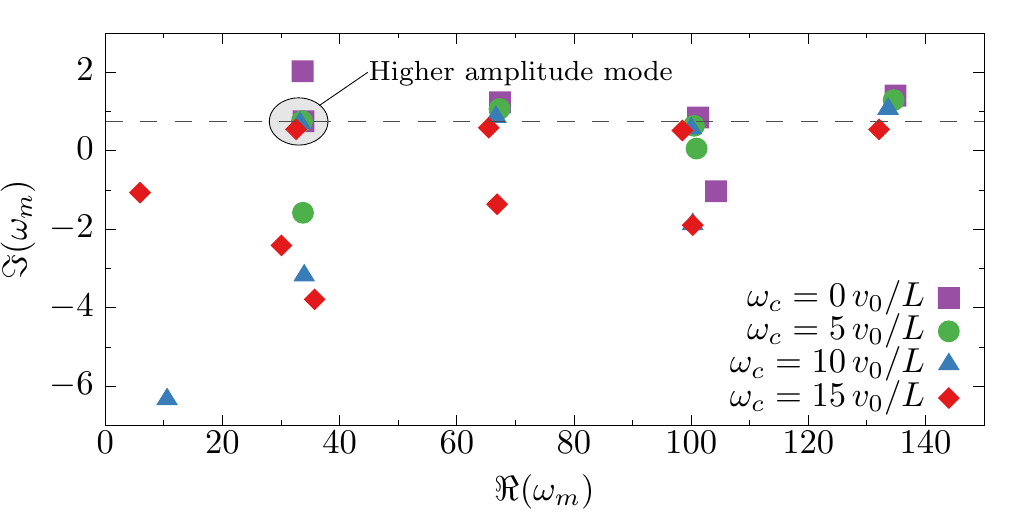}
    \caption{Higher order dynamic mode decomposition frequencies, $\Re(\omega_m)$, and growth rates, $\Im(\omega_m)$ (in units of $v_0/L$), the modes with higher amplitude are displayed with stronger color. Dashed line marking the theoretical growth rate from \eqref{eq:omegatotal}. The decomposition was obtained from the linear regime ($t\lesssim 2 \,L/v_0$) of the average drain-to-source current for different values of cyclotron frequency $\omega_c$ with $S=20v_0$ and $v_F=10v_0$.}
    \label{fig:HoDMD}
\end{figure}

To further analyze and quantify the impact of $\omega_c$ on the electronic fluid, the numerical results were evaluated with higher order dynamic mode decomposition (HODMD) \cite{LeClainche2017HigherDecomposition} resorting to PyDMD software \cite{Demo2018PyDMD:Decomposition}. The direct outputs of the fluid equations have been firstly integrated to obtain the average drain-to-source current; this enables the analysis to be performed on a lower dimensionality quantity that retains the dynamic of the system. Then, the HODMD algorithm was applied to the linear growth portion of the signal, i.e. before the nonlinear saturation effects, which corresponds to $t\lesssim 1.5L/v_0$. Although HODMD can perfectly deal with the transition to the saturation regime, the eigenmodes and complex frequencies thus retrieved do not necessarily reflect the values predicted by linear theory. Figure \ref{fig:HoDMD} shows an example of such results where the overall decrease of growth rate is evident, with the growth rates from the $\omega_c=0$ case exceeding the subsequent results with magnetic field. Moreover, the predicted slight drift of the main frequency towards higher values can also be observed.   

\section{Conclusions}

The theoretical study of electronic transport in graphene is a challenging task, covering several regimes and interactions, and resorting to complex techniques. Nonetheless, the hydrodynamic models provide a semi-classical description capable of recovering the behavior and properties of such quantum fluids while also allowing numerical simulation with well-established methods. 
However, it is vital to stress that conventional fluid equations --- for instance, the Cauchy momentum equation --- can not be bluntly applied and that the variation of the effective mass with the numerical density introduces a correction in the nonlinear convective term, breaking the symmetry of the dispersion relation in the presence of a base drift of the fluid.
 
The presented model evince that the presence of a weak transverse magnetic field dramatically changes the nature of the plasmons for small $k$, opening a gap in the dispersion relation, imposing a cut-off on the feasible frequencies of such systems. Furthermore, our numerical results point out that the magnetic field impairs the growth of the DS instability, a result that, to our knowledge, has not yet been reported in this context. Such reduction of the growth rate is practically unnoticeable for the deep subsonic flows on which technological applications are bound to operate. Yet, the frequency itself can be increased for moderate values of Mach number before reaching the gap cut-off. Moreover, our results suggest that the DS configuration in a magnetized FET has the potential to function as a directional coupler operating in the THz regime \cite{He2014Graphene-basedCoupler}.   
In future studies, other magnetic effects could be addressed, either with DS mechanism or exploring other instability processes. Namely, drift instabilities considering the enhanced diamagnetic drift arising from the gated scenario. Lastly, the presence of magnetic field would also lead to the emergence of an odd viscosity \cite{Avron1998OddViscosity} contribution with potentially interesting effects, such as topologically protected edge states and new exotic dynamics.

\begin{acknowledgments}

The authors acknowledge the funding provided by Funda\c{c}\~ao para a Ci\^encia e a Tecnologia (FCT-Portugal) through the Grant No. PD/BD/150415/2019 and the Contract No. CEECIND/00401/2018.

\end{acknowledgments}

\section*{AIP Publishing Data Sharing Policy}
The data that support the findings of this study are available from the corresponding author upon reasonable request.

\bibliography{References}
\end{document}